\documentclass[sigconf,timestamp=true]{acmart}

\usepackage{booktabs} 
 \renewcommand\footnotetextcopyrightpermission[1]{} 
\setcopyright{none}

\usepackage{url}
\usepackage{mathptmx}
\usepackage{amsmath}
\usepackage{times}
\usepackage{xcolor}
\usepackage{amsmath}
\usepackage[utf8]{inputenc}
\usepackage{epsfig,endnotes,color}
\usepackage{array}
\usepackage{multirow}
\usepackage{xspace}
\usepackage{verbatim}
\usepackage{wrapfig}
\usepackage{mathptmx}
\usepackage[english]{babel}
\usepackage{graphicx}
\usepackage{csvsimple}
\usepackage{epstopdf}
\usepackage{enumitem}
\usepackage{datetime}
\usepackage{cleveref}

\usepackage{amssymb}
\usepackage{tabularx}
\usepackage[linesnumbered,ruled,vlined]{algorithm2e}
\usepackage{float}
 
\begin{document}
\title{Differential Privacy By Sampling}

\author{Joshua Joy, Mario Gerla}

\affiliation{    \institution{University of California - Los Angeles}
}
\email{{jjoy,gerla}@cs.ucla.edu}

\newcommand{\titlename}{Sampling Privacy~}

\begin{abstract}

In this paper we present the \titlename mechanism for privately releasing personal data. \titlename is a sampling based privacy mechanism that satisfies differential privacy.

\end{abstract}

\maketitle

\section{Introduction}

Publicly releasing data with exact answers to queries (without sanitization) has incurred numerous privacy violations and attacks relating to unintentional medical data disclosure of high profile governors~\cite{Sweene02}, shutdowns of seemingly innocuous open data machine learning competitions~\cite{netflix-privacy-lawsuit}, and unintentional sharing of mobility patterns of high profile US citizens with foreign governments~\cite{uber-privacy-china}. k-anonymity introduced by Sweeney in 1998~\cite{Sweene02} was among the first privacy techniques to address publicly releasing data in a privacy-preserving manner. Roughly speaking, k-anonymity seeks to blend a single data owner's personal attribute with at least $k$ other data owners such that the single data owner is indistinguishable from $k-1$ other data owners. For example, if a particular data owner's record reporting a particular disease is publicly released with $1000$ other data owners records with the same disease, the data owner is indistinguishable from $999$ other data owners.

However, there are known impossibility results for attempts to preserve privacy while releasing exact answers. Dinnur and Nissim showed in 2003 that it is impossible to reveal exact aggregation information while simultaneously preserving privacy (against a polynomial adversary)~\cite{DBLP:conf/pods/DinurN03}. Thus, perturbation \textit{must} be injected in order to guarantee privacy (privacy defined as an adversary is unable to determine the value of a targeted individual with a probability greater than 50\%)~\cite{DBLP:conf/pods/DinurN03}. The implication is there will be some notion of absolute error due to the required perturbation.

In this paper, our goal is to achieve the notion of \textit{scalable} privacy. That is, as the population increases the privacy should \textit{strengthen}. Additionally, the absolute error should remain at worst \textit{constant}. For example, suppose we are interested in understanding a link between eating red meat and heart disease. We start by querying a small population of say 100 and ask ``Do you eat red meat and have heart disease?". Suppose 85 truthfully respond ``Yes". If we know that someone participated in this particular study, we can reasonably infer they eat red meat and have heart disease regardless the answer. Thus, it is difficult to maintain privacy when the majority of the population truthfully responds ``Yes". 

Querying a larger and diverse population would protect the data owners that eat red meat and have heart disease. Let's say we query a population of 100,000 and it turns out that 99.9\% of the population is vegetarian. In this case, the vegetarians blend with and provide privacy protection of the red meat eaters. However, we must be careful when performing estimation of a minority population to ensure the sampling error does not destroy the underlying estimate.

Differential privacy is one such privacy definition which captures this concept of perturbation while preserving accuracy and has emerged as the gold standard~\cite{DBLP:conf/icalp/Dwork06,DBLP:conf/tcc/DworkMNS06}. Roughly speaking, differential privacy says that a query against two databases which differ by at most one row is $\epsilon$ indistinguishable (where $\epsilon$ is the acceptable privacy leakage). Thus, a data owner is able to safely participate in sharing their personal data, as there is minimal risk that personal information will be leaked.

A popular technique which satisfies differential privacy is the randomized response mechanism, originally proposed in the 1960s~\cite{warner1965randomized,fox1986randomized}. Randomized response is used by many companies today (e.g., Apple, Google~\cite{DBLP:conf/ccs/ErlingssonPK14}) due to its simplicity while satisfying the differential privacy guarantee. 

The randomized response mechanism perturbs the truthful answer by having the data owner flip two biased coins (e.g., two independent Bernoulli trials)~\cite{warner1965randomized,fox1986randomized}. The randomized response mechanism proceeds as follows. The data owner flips the first coin. If the first coin comes up heads, answer truthfully. Else flip the second coin and answer ``Yes" if heads and ``No" if tails. The perturbation is performed by the  data owner (induced by the coin flips), as opposed to trusting a centralized service to perturb the truthful response. Estimation occurs by aggregating the privatized responses, subtracting the expected value of the second coin toss, and then finally dividing by the probability of the first coin toss. For example, suppose the coin tosses are configured independently with $flip_1=0.85$, $flip_2=0.3$ and $100$ data owners. The coin toss parameters are published publicly while the number of data owners is private and needs to be estimated. We estimate by aggregating the privatized responses from all data owners, subtracting the expected value of $(1-0.85) \times 0.3 \times 100$ and dividing by $0.8$ \footnote{For instance with a 60\% truthful population, the answer to the first toss is $0.6 \times 0.85=0.51$ and the answer to the second toss is $(1-0.85) \times 0.3=0.045$}.

However, a drawback to the randomized response mechanism is that the estimation error quickly increases with the population size due to the underlying truthful distribution distortion. For example, say we are interested in how many vehicles are at a popular stretch of the highway. Say we configure $flip_1=0.85$ and $flip_2=0.3$. We query 10,000 vehicles asking for their current location and only 100 vehicles are at the particular area we are interested in (i.e., 1\% of the population truthfully responds ``Yes"). The standard deviation due to the privacy noise will be 21 \footnote{($\sqrt{(1-0.85) \times 0.3 \times 10,000}$)} which is slightly tolerable. However, a query over one million vehicles (now only 0.01\% of the population truthfully responds ``Yes") will incur a standard deviation of 212. The estimate of the ground truth (100) will incur a \textit{large} absolute error when the aggregated privatized responses are two or even three standard deviations (i.e., 95\% or 99\% of the time) away from the expected value, as the mechanism subtracts \textit{only} the expected value of the noise.

Some protocols which leverage the randomized response mechanism have made assumptions that the underlying truthful population is fixed at 2/3 or 3/4 of the total population in order to preserve accuracy. However, it's not clear what privacy guarantees will be provided as any adversary is able to successfully guess with greater than 50\% probability the value of any data owner in such a population. For example, suppose our query is how many home owners reside within 15 blocks from the beach, yet we ask only those home owners within 20 blocks from the beach.

The Laplace mechanism was introduced as a way to add privacy noise independent of the database size~\cite{DBLP:conf/tcc/DworkMNS06} by drawing privacy noise from the Laplace distribution. The Laplace mechanism is calibrated to the max difference between any two rows in the database. That is, the noise is sufficient to protect the max leakage that any particular data owner induces. For example, first a service aggregates all the data owners truthful responses. Then, the service draws from the Laplace distribution by calibrating the variance according to the desired privacy strength. Drawing from other distributions such as Gaussian also satisfies differential privacy, though the Laplace mechanism is preferred as it's mathematically cleaner~\cite{DBLP:journals/fttcs/DworkR14}.

However, there is a drawback to the Laplace mechanism in graph datasets such as social networks~\cite{DBLP:conf/tcc/GehrkeLP11,DBLP:conf/crypto/GehrkeHLP12} or vehicle commuting patterns. Even if a particular data owner does \textit{not} participate, their friends that do participate leak information that can be used to deprivatize the targeted data owner (e.g., shadow profiles). For example, it is possible to learn political beliefs or sexual orientation even if a particular individual does not participate and maintain an active profile in an online social network. An adversary simply needs to analyze the similarity metrics amongst the social circles that a data owner participates in to understand politics beliefs or sexual orientation~\cite{DBLP:conf/cosn/SarigolGS14,10.1371/journal.pone.0034740,facebook-friends,Kosinski_Stillwell_Graepel_2013,FM2611}.

Furthermore, if the graph structures of the social network are eventually anonymized and released, an adversary simply needs to participate and influence the graph structure (e.g., joining a social network) to learn and influence the actual social graph before it's privatized and released. Thus, there needs to be a mechanism which also perturbs the underlying \textit{structure} of the data itself and preserves accuracy as the underlying distribution structure becomes distorted.

Sampling whereby responses are randomly discarded reduces the the graph dependencies leaked by a targeted individuals connections. The severed connections reduces the social circle size and makes it challenging for the adversary to make similarity inferences from reduced social circles alone. Thus, it has been shown that the strength of privacy mechanisms are increased by applying sampling and reducing the privacy leakage~\cite{DBLP:conf/stoc/NissimRS07,DBLP:conf/focs/KasiviswanathanLNRS08,DBLP:conf/ccs/LiQS12,DBLP:conf/crypto/GehrkeHLP12}. 

However, we are interested in more than extending existing privacy mechanisms. We ask can we achieve privacy by sampling alone? Can sampling based plausible deniability be provided while simultaneously maintaining constant error as the population increases and the underlying truthful population becomes distorted?

In this paper, we present the \titlename mechanism. \titlename is a distributed sampling approach in which a data owner answers with (contradictory) responses yet maintains constant absolute error as the population increases and the truthful distribution becomes distorted. \section{Related Work}

~\\ \noindent \textbf{Privacy Definitions.} Differential privacy~\cite{DBLP:conf/icalp/Dwork06,DBLP:conf/tcc/DworkMNS06,DBLP:conf/eurocrypt/DworkKMMN06,DBLP:journals/fttcs/DworkR14} has been proposed as a privacy definition such that anything that can be learned if a particular data owner is added to the database could have also been learned before the data owner was added. A data owner is thus ``safe" to participate as statistical inferences amongst the aggregate are learned yet specific information regarding the individual is not learned.	

Zero-knowledge privacy~\cite{DBLP:conf/tcc/GehrkeLP11} is a cryptographically influenced privacy definition that is strictly stronger than differential privacy. Crowd-blending privacy~\cite{DBLP:conf/crypto/GehrkeHLP12} is weaker than differential privacy; however, with a pre-sampling step, satisfies both differential privacy and zero-knowledge privacy. However, these mechanisms are suited for the centralized system model and rely on aggressive sampling, which significant degrades the accuracy estimations.

Distributional privacy~\cite{DBLP:journals/jacm/BlumLR13} is a privacy mechanism which says that the released aggregate information only reveals the underlying ground truth distribution and nothing morre. Each data owner is protected by the randomness of the other randomly selected data owners rather than by adding explicit privacy noise to the output. The indistinguishability from the underlying distribution protects individual data owners and is strictly stronger than differential privacy. However, it is computationally inefficient though can work over a large class of queries known as Vapnik-Chervonenkis (VC) dimension.

~\\ \noindent \textbf{Sampling.} Sampling whereby a centralized aggregator randomly discards responses has been previously formulated as a mechanism to amplify privacy~\cite{DBLP:conf/crypto/ChaudhuriM06,DBLP:conf/stoc/NissimRS07,DBLP:conf/focs/KasiviswanathanLNRS08,DBLP:conf/ccs/LiQS12,DBLP:conf/crypto/GehrkeHLP12}. The intuition is that when sampling approximates the original aggregate information, an attacker is unable to distinguish when sampling is performed and which data owners are sampled. These privacy mechanisms range from sampling without a sanitization mechanism, sampling to amplify a differentially private mechanism, sampling that tolerates a bias, and even sampling a weaker privacy notion such as k-anonymity to amplify the privacy guarantees. 

However, sampling alone has several issues. First, data owners are not protected by plausible deniability as data owners do not respond ``No". Second, the estimation of the underlying truthful ``Yes" responses quickly degrades as we increase the population that truthfully responds ``No".

~\\ \noindent \textbf{Multi-party Computation.} Multi-party computation (MPC) is a secure computation model whereby parties jointly compute a function such that each party only learns the aggregate output and nothing more. However, MPC mechanisms that release the \textit{exact} answer has no strong privacy guarantees against active privacy attacks, particularly when the data is publicly published. A participant that does not perturb their responses and provides their \textit{exact} answer is easily attacked by an adversary that knows the values of $n-1$ participants. For example, an adversary first runs a counting query that includes all $n$ data owners and then runs a second counting query over $n-1$ data owners (the targeted data owner is the excluded row). Subtracting the two results \textit{reveals} the value of the targeted data owner. In contrast, the differential privacy model assumes a strong adversary that knows the $n-1$ data owner values. There are attempts to combine MPC and differential privacy, though these are outside the scope of this paper. \section{Toy Construction}
\label{sec:toyconstruction}

Let us first consider a simplistic sampling based privacy approach and show that it has limitations, in particular it is not scalable as it does \textit{not} maintain constant error as the population increases. This will motivate the introduction of a more sophisticated sampling based scheme -  the \titlename mechanism.

Consider a population whereby only 5\% of the population satisfies a particular attribute (for example at a particular location). These 5\% of the population should truthfully respond ``Yes" while the remaining 95\% subpopulation truthfully responds ``No". We refer to each subpopulation as $Yes_{pop}$ and $No_{pop}$ respectively. 

It should be noted that the $Yes_{pop}$ is fixed. That is, at a given time instance the number of people physically at a particular location is physically constrained. The query population is adjustable by including those \textit{not} at a particular location. Thus, only the $No_{pop}$ increases. Say, instead of querying only Manhattan vehicles about presence in Times Square, we query all vehicles in New York State. One approach is to grow the query population size by several orders of magnitude.

For example, let's ask a counting query of how many people are currently at the Statue of Liberty. We then privately release the result of this counting query privately to protect the personal data of each data owner.

We could query only those physically at the Statue of Liberty ($Yes_{pop}$). However, if we know that any particular data owner participated in the query we know absolutely certain their location is the Statue of Liberty. Thus, the privacy protection is quite limited regardless of any perturbation performed for this particular query. 

The privacy protection would be \textit{strengthened} if we query \textit{everyone} in New York State. The additional data owners provide plausible deniability by increasing the potential pool of candidates that sometimes respond ``Yes" indicating they are at the Statue of Liberty. Now, if we know that someone participated in the query all we can immediately deduce is they are ``somewhere" in New York State.

We now illustrate a toy construction of our sampling based privacy scheme, showing that it overcomes simple sampling drawbacks. Our goal is to increase the pool of candidates (e.g., in our example those currently \textit{not} at the Statue of Liberty) while simultaneously maintaining constant error.

\subsection{Threat Model}
\label{sec:threatmodel}
The attack:  an adversary can utilize the database size (number of participants) to deduce if a particular individual is included. However, the \textit{exact} population (database) size or \textit{exact} number of participating data owners is not published or released. This mitigates auxiliary attacks whereby the adversary uses the exact counts to reconstruct the database.

The attack: an adversary can individually inspect the responses of each data owner to ascertain their truthful response. However, we select sampling probabilities less than 50\% so that an adversary does not gain an inference advantage of greater than 50\%. We also require a distributed set of aggregators or trusted aggregator whereby at least one aggregator does not collude with the others.

\subsection{Sampling and Noise}
\label{sec:samplingnoise}

Our first question is how to construct a mechanism such that those that truthfully respond ``No" to sometimes respond ``Yes"? We could leverage the $No_{pop}$ population by having a subpopulation respond ``Yes''. To perform the final estimation we need to subtract the estimated added noise. However, let us show more formally that sampling and random noise alone does not work as there are more unknown variables than equations.

~ \\
\noindent \textbf{(Sampling and Noise Response)} Each data owner privatizes their truthful value by performing the following Bernoulli trial. Let $\pi_{s}$ be the sampling probability .

\begin{equation}
  Privatized~Value =
  \begin{cases}
    1 & \text{with probability $\pi_{s}$} \\
    \varnothing & \text{with probability $1-\pi_{s}$} \\
    0 & never
  \end{cases}
\end{equation}

That is, each data owner is sampled uniformly at random to respond ``Yes". A sampled portion of the $Yes_{pop}$ truthfully responds ``Yes". Privacy noise is provided by the sampled portion of the $No_{pop}$ which responds ``Yes". The remaining population does not participate and there is not a single data owner which responds ``No". 

It is not possible to estimate the underlying $Yes_{pop}$ as our estimator is an unsolvable system of equations with one equation (the aggregated privatized count) and two unknowns ($Yes_{pop}$ and $No_{pop}$). The $Yes_{pop}$ must somehow become distinguishable from the $No_{pop}$ in order to estimate the underlying population of those that truthfully should respond ``Yes". We also observe that there must be some privacy leakage in order to have any notion of accuracy when the underlying data is perturbed.

\subsection{Sampling and Plausible Deniability}

After the previous false starts, we now attempt to construct a mechanism for which we can estimate the underlying population and for which plausible deniability is provided. That is, data owners occassionally respond \textit{opposite} to their truthful response based on the resulting coin tosses.

~ \\
\noindent \textbf{(Sampling and Plausible Deniability Response)} We now describe how to achieve both sampling and plausible deniability below. There are two separate protocols. One for those that truthfully respond ``Yes" . Conversely, another for those that truthfully respond ``No". The protocols are named Privatized $Value_{Yes}$ and Privatized $Value_{No}$ respectively. The protocols are defined below.

~ \\
\noindent \textbf{(Privatization)} Let $\pi_{s}$ be the sampling probability that determines whether a data owner with data participates or not. We also use two independent and biased coins. $\pi_1$ and $\pi_2$ refer to the first and second biased coin toss respectively.

\begin{equation}
  Privatized~Value_{Yes} =
  \begin{cases}
    \varnothing & \text{with probability $1-\pi_{s}$} \\
    1 & \text{with probability } \\
    ~ & \text{$\pi_{s} \times (\pi_1 + (1-\pi_1) \times \pi_2) $} \\
    0 & \text{otherwise}
  \end{cases}
\end{equation}

That is, a data owner responds ``Yes" with probability $\pi_{s} \times (\pi_1 + (1-\pi_1) \times \pi_2) $ from the $Yes_{pop}$ subpopulation.  They do not participate with probability $1-\pi_{s}$. Otherwise they respond ``No".

\begin{equation}
  Privatized~Value_{No} =
  \begin{cases}
    \varnothing & \text{with probability $1-\pi_{s}$} \\
    1 & \text{with probability} \\
    ~ &  \text{$\pi_{s} \times ((1-\pi_1) \times \pi_2) $} \\
    0 & \text{otherwise}
  \end{cases}
\end{equation}

That is, a data owner responds ``Yes" with probability $\pi_{s} \times ((1-\pi_1) \times \pi_2) $ from the $No_{pop}$ subpopulation. They do not participate with probability $1-\pi_{s}$. Otherwise they respond ``No".

We have plausible deniability protection as each data owner will occasionally respond opposite to their truthful response. The total amount of privacy noise is generated with probability $\pi_{s} \times ((1-\pi_1) \times \pi_2)$ from each subpopulation and needs to be removed in order to perform estimation of the underlying truthful population as we show below.

~ \\
\noindent \textbf{(Expected Values)} We now formulate the expected values in order to carry out the estimation. The expected value of those that respond `1' (i.e., privatized ``Yes") is the sum of the binomial distribution of each subpopulation.

\begin{multline}
E[1] = \pi_{s} \times \pi_1 \times Yes_{pop} + \\ \pi_{s} \times (1-\pi_1) \times \pi_2 \times (Yes_{pop} + No_{pop}) 
\end{multline}

~ \\
\noindent \textbf{(Estimator)} We solve for $Yes_{pop}$ by the following. Let the aggregated privatized counts be denoted as $Private~Sum$. 
\begin{multline}
\label{eq:rr}
Yes_{pop} = \frac{Private~Sum - \pi_{s} \times (1-\pi_1) \times \pi_2 \times (Yes_{pop} + No_{pop})}{\pi_{s} \times \pi_1}
\end{multline}

That is, we first subtract the expected value of the privacy noise. We then divide by the sampling probability $\pi_{s}$ which determines if a data owner with data participates. $\pi_1$ is the sampling parameter which determines how frequently a data owner truthfully responds ``Yes" from the  $Yes_{pop}$ subpopulation. It should be noted that Equation~\ref{eq:rr} is precisely the structure of the randomized response mechanism~\cite{warner1965randomized,fox1986randomized}.

Examining the structure of  Equation~\ref{eq:rr} we make the following observation. Increasing the $No_{pop}$ population and correspondingly reducing the underlying truthful percentage of the population to say 5\% and below will induce large sampling error. Unfortunately, as the $Yes_{pop}$ is physically constrained and fixed (see Section~\ref{sec:toyconstruction}), it is the only population that can be scaled up.

We desire better calibration over the privacy mechanism and a mechanism which maintains constant error as the $No_{pop}$ population scales up. We now introduce the \titlename mechanism. \section{\titlename Mechanism}
\label{sec:haystackmechanism}

We now describe the \titlename mechanism that achieves constant error even as the population which truthfully responds ``No" ($No_{pop}$) increases. We motivate our example using location coordinates, though the mechanism applies to all real valued data. We describe the general mechanism and then formally describe it below.

\begin{figure}[t!]
    \centering
    \includegraphics[width=0.65\columnwidth]{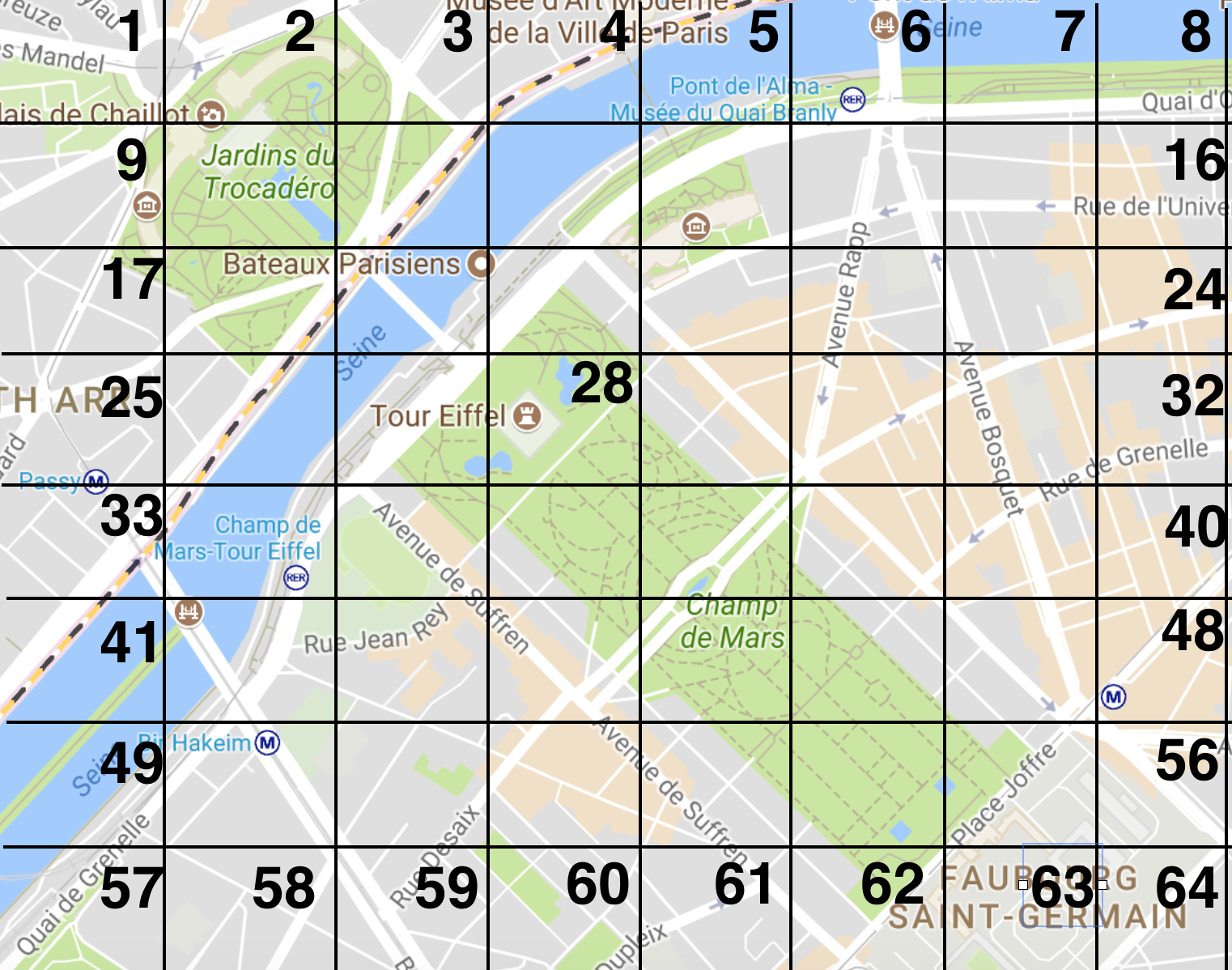}
    \caption{\textbf{(Location Discretization)} Each location coordinate (latitude,longitude) is discretized to a location identifier.}
        \label{fig:discretization}
\end{figure}

Suppose a data owner currently at the Eiffel Tower participates in the protocol. First, the location is discretized to a location identifier (ID) as seen in Figure~\ref{fig:discretization}. For example, using a 32 bit identifier provides 4 billion possible locations, which covers a 9,500 x 9,500 mile square with 0.15 mile sections for a total of 90 million square miles. For comparison Paris is 41 square miles, London is 607 square miles, New York City is 305 square miles and Beijing is 6,336 square miles~\cite{us-cities-area}. In Figure~\ref{fig:discretization} the Eiffel Tower corresponds to location ID 28.

Then, suppose the data owner at the Eiffel Tower is sampled and selected. In the first round the selected data owner should respond ``No". The remaining data owners uniformly at random respond either ``Yes" or ``No" regardless of their truthful response. The counts of ``Yes" and ``No" are aggregated.

In the second round the selected data owner should respond ``Yes". The remaining data owners remain with their response in the first round. The counts of ``Yes" and ``No" are aggregated.

Now, by subtracting the ``Yes" counts of round one from round two we can compute the sampled population count. Finally, dividing the sampled population count by the sampling parameter computes the estimated number of data owners currently at the Eiffel Tower.

There are three privacy points to observe. First, a majority of the population provides privacy noise by uniformly at random responding either ``Yes" or ``No" regardless of their truthful response. Second, plausible deniability is provided as each data owner probabilistically responds opposite of their truthful response. Finally, \textit{every} data owner acts as a candidate for the truthful population. Our assumption is that every data owner is active in both rounds and only the aggregate counts are released.

We now formally describe the \titlename mechanism.

\subsection{Binary Value}
\label{sec:binaryvalue}

We now formally describe the binary version of the \titlename mechanism whereby data owners respond either ``No" or ``Yes", either $0$ and $1$, respectively. Following from the Eiffel Tower example above, $0$ refers to currently \textit{not} at the Eiffel Tower and $1$ refers to currently at the Eiffel Tower.

~ \\
\noindent \textbf{(Round One)} In the first round, the data owners flip a three sided die. The three probabilities are $\pi_{0}$, $\pi_{s}$, and $1-\pi_{0}-\pi_{s}$ where $\pi_{0}$ refers to the probability of responding to the output $0$, $\pi_{s}$ is the sampling parameter which also responds to the output $0$, and otherwise $1-\pi_{0}-\pi_{s}$ is the probability to respond to the output $1$. The first round establishes the baseline used to estimate the count by having those with the sampling probability assigned to output $0$ corresponding to ``No".

\begin{equation}
\label{eq:roundonebinary}
  Round~One =
  \begin{cases}
    0 & \text{with probability $\pi_{0}$} \\
    0 & \text{with probability $\pi_{s}$} \\
    1 & \text{with probability $1-\pi_{0}-\pi_{s}$}
\end{cases}
\end{equation}

~ \\
\noindent \textbf{(Round Two)} In the second round, we \textit{shift} the sampled population that truthfully respond ``Yes" from $0$ to $1$. This shift of the subpopulation allows us to estimate the underlying distribution as we will see below.

\begin{equation}
\label{eq:roundtwoyesbinary}
  Round~Two_{Yes} =
  \begin{cases}
    0 & \text{with probability $\pi_{0}$} \\
    \mathbf{1} & \textbf{with probability $\pi_{s}$} \\
    1 & \text{with probability $1-\pi_{0}-\pi_{s}$}
\end{cases}
\end{equation}

That is, the truthful ``Yes" responses that responded ``No" in the first round now respond ``Yes" in the second round.

The sampled population that truthfully respond ``No" stays at $0$. The remaining data owners remain at their output chosen in round one.

\begin{equation}
  Round~Two_{No} =
  \begin{cases}
    0 & \text{with probability $\pi_{0}$} \\
    0 & \text{with probability $\pi_{s}$} \\
    1 & \text{with probability $1-\pi_{0}-\pi_{s}$}
\end{cases}
\end{equation}

The separate protocols for each subpopulation allows us to increase the $No_{pop}$ while retaining constant error.

~ \\
\noindent \textbf{(Expected Values)} We now formulate the expected values as follows. The subscript refers to the round number. That is, $0_{1}$ refers to output 0, round 1. The first round of expected values for each output are:

\begin{align}
\begin{split}
\label{eq:roundonesinglequery}
E[0_{1}] & = \pi_{0} \times \mathit{TOTAL_{pop}} + \pi_{s} \times \mathit{TOTAL_{pop}} \\
E[1_{1}] & = (1-\pi_{0}-\pi_{s}) \times \mathit{TOTAL_{pop}}
\end{split}
\end{align}

The second round of expected values for each output are:

\begin{align}
\begin{split}
\label{eq:roundtwosinglequery}
E[0_{2}] & = \pi_{0} \times \mathit{TOTAL_{pop}} + \pi_{s} \times No_{pop}\\
E[1_{2}] & = (1-\pi_{0}-\pi_{s}) \times \mathit{TOTAL_{pop}} + \pi_{s} \times \mathit{Yes_{pop}}
\end{split}
\end{align}

~ \\
\noindent \textbf{(Estimator)} We solve for the $Yes_{pop}$ population by subtracting round two by round one as follows.  Let $Private~Sum~0_{1}$ refer to the aggregated privatized counts for output 0, round 1.

\begin{align}
\begin{split}
Yes_{pop} & = \frac{Private~Sum~1_{2} - Private~Sum~1_{1}}{\pi_{s}}
\end{split}
\end{align}

That is, we recover the sampled population by subtracting the privatized sum of the shifted output 1, round 1 from round 2 (Equations ~\ref{eq:roundonebinary} and ~\ref{eq:roundtwoyesbinary}). Then we obtain the estimate by dividing by the sampling parameter.

This mechanism constrains the sampling error to solely the $Yes_{pop}$ as seen by Equations~\ref{eq:roundonesinglequery} and ~\ref{eq:roundtwosinglequery}. We are able to scale the $No_{pop}$ yet retain constant error. Plausible deniability is provided as each data owner may respond ``Yes" or ``No" based on the coin toss parameters.

\subsection{Multiple Values}
\label{sec:multiplevalues}

We now examine how to privatize the multiple choice scenario whereby there are multiple values and the data owner should select a single value. We extend the binary value mechanism defined in the previous section. Multiple values are applicable to most real-world scenarios (as opposed to the binary value mechanism). The location coordinate grid scenario, explained in Section~\ref{sec:haystackmechanism} and illustrated in Figure~\ref{fig:discretization}, explains a scenario where there are multiple locations (i.e., location IDs) and the data owner is currently at a single location ID. Recall that the data owner's truthful response is discretized to an integer value greater than 0.

The number of values of the outputs should equal to the desired number of values \textit{plus} an output (used for baseline calibration in the first round). For example, if there are 9 locations to monitor, there should be a total of 10 output values, where 0 is the additional output.

Let there be a total of $V$ possible discretized output values (e.g., $V$ possible locations). The total number of output values should then be $V+2$.  Let $V'$ be the truthful output value for a particular data owner.

~ \\
\noindent \textbf{(Round One)} The first round is similar to the first round of the binary value mechanism. The data owner tosses a $V+1$ sided die with probabilities $\pi_{0}$, $\pi_{1}$,...,$\pi_{V}$, and $\pi_{s}$. That is, each data owner uniformly at random selects an output independent of their truthful value. However, a small sampling fraction writes to the output 0 (in order to baseline the values used for estimation). 

\begin{equation}
  Round~One =
  \begin{cases}
    0 & \text{with probability $\pi_{0}$} \\
    0 & \text{with probability $\pi_{s}$} \\
    1 & \text{with probability $\pi_{1}$} \\
    2 & \text{with probability $\pi_{2}$} \\
    ... \\
    V & \text{with probability} \\
    		& \text{$1-\pi_{0}-\pi_{1}-...-\pi_{V}-\pi_{s}$}
\end{cases}
\end{equation}

~ \\
\noindent \textbf{(Round Two)} In the second round, the sampled data owners that flipped $\pi_{s}$ now choose their truthful output value $V'$. The remaining data owners stay at the output chosen in round one.

\begin{equation}
  Round~Two =
  \begin{cases}
    0 & \text{with probability $\pi_{0}$} \\
    1 & \text{with probability $\pi_{1}$} \\
    2 & \text{with probability $\pi_{2}$} \\
    ... \\
    V' & \text{with probability $\pi_{v'}$} \\
    \mathbf{V'} & \textbf{with probability $\pi_{s}$} \\
    ... \\
    V & \text{with probability} \\
    		& \text{$1-\pi_{0}-\pi_{1}-...-\pi_{V}-\pi_{s}$}
\end{cases}
\end{equation}

This step allows us to estimate the underlying truthful distribution.

~ \\
\noindent \textbf{(Expected Values)} The expected values are as follows.

The \textbf{first round} of expected values are as follows. The output 0 includes the fraction of the population which is randomly assigned to output 0. Additionally, the sampled populations of those that answer ``Yes" truthfully at each output are initially assigned to output 0 in the first round. This serves as the baseline to perform estimation as we will see after the second round.

\begin{align}
\begin{split}
E[0_{1}] & = \pi_{0} \times \mathit{TOTAL} + \pi_{s} \times \sum_{n=1}^{V} Yes_{pop},{n} \\
E[1_{1}] & = \pi_{1} \times \mathit{TOTAL} \\
E[2_{1}] & = \pi_{2} \times \mathit{TOTAL} \\
... \\
E[V_{1}] & = \pi_{V} \times \mathit{TOTAL}
\end{split}
\end{align}

The \textbf{second round} of expected values are as follows. The sampled population is assigned to their truthful output $V'$. The remaining data owners stay at the output assigned in round one. 

\begin{align}
\begin{split}
E[0_{2}] & = \pi_{0} \times \mathit{TOTAL} \\
E[1_{2}] & = \pi_{1} \times \mathit{TOTAL} \\
E[2_{2}] & = \pi_{2} \times \mathit{TOTAL} \\
... \\
E[V'_{2}] & = \pi_{V'} \times \mathit{TOTAL} + \pi_{s} \times \mathit{Yes_{pop}} \\
... \\
E[V_{2}] & = \pi_{V} \times \mathit{TOTAL}
\end{split}
\end{align}

~ \\
\noindent \textbf{(Estimator)} We iterate over each output $1...V$ and solve for each output $V'$ by subtracting round two by round one as follows:

\begin{align}
\begin{split}
\mathit{Yes_{pop},V'} & = \frac{Private~Sum~V'_{2} - Private~Sum~V'_{1}}{\pi_{s}}
\end{split}
\end{align}

\subsection{Privacy Guarantee}
\label{sec:dpprivacyguarantee}

The \titlename mechanism satisfies differential privacy as we show in this section. We first examine the binary value mechanism and then the multiple values mechanism.

~ \\
\noindent \textbf{Differential Privacy.} Differential privacy has become the \emph{gold standard} privacy mechanism which ensures that the output of a sanitization mechanism does not violate the privacy of any individual inputs.  

\begin{definition}[\cite{DBLP:conf/icalp/Dwork06,DBLP:conf/tcc/DworkMNS06}]{($\epsilon$-Differential Privacy).}
\label{def:differentialprivacy}
A privacy mechanism $San()$ provides $\epsilon$-differential privacy if, for all datasets $D_1$ and $D_2$ differing on at most one record (i.e., the Hamming distance $H()$ is $H(D_1,D_2) \leq 1$), and for all outputs $O \subseteq Range(San())$:
\begin{equation}
\sup_{D_1,D_2}\frac{\Pr[San(D_1) \in O]}{\Pr[San(D_2) \in O]} \leq exp(\epsilon)
\label{eqn:dp}
\end{equation}
\end{definition}

That is, the probability that a privacy mechanism $San$ produces a given output is almost independent of the presence or absence of any individual record in the dataset.  The closer the distributions are (i.e., smaller $\epsilon$), the stronger the privacy guarantees become and vice versa. That is, a larger $\epsilon$ means that the two dataset distribution are far apart and leaks more information. A single record will induce distinguishable output fluctuations. We desire smaller $\epsilon$ values to induce $\epsilon$ \textit{indistinguishability}.

\subsection{Differential Privacy Guarantee}
~ \\
\noindent \textbf{(Binary Value)} The differential privacy leakage is measured as the maximum ratio of the binary output given the underlying truthful answer is ``Yes" and ``No" respectively.

In round one, there is no privacy leakage as both output 0 and 1 are both equally likely and indistinguishable given the truthful answer is either ``No" or ``Yes" respectively.  Thus, we analyze the privacy leakage in round two. 

For \textbf{output 0, round 2}  ($0_{2}$) the privacy leakage is as follows:

\begin{equation}
\epsilon_{DP} = \max \Bigg({\ln \bigg(\frac{\Pr[0_{2} | ``Yes"]} {\Pr[0_{2} | ``No"]} \bigg),\ln \bigg(\frac{\Pr[0_{2} | ``No"]} {\Pr[0_{2} | ``Yes"]} \bigg)} \Bigg)
\end{equation}

\begin{equation}
\frac{\Pr[0_{2} | ``Yes"]} {\Pr[0_{2} | ``No"]} = \frac{ \pi_{0} + \pi_{s}}{ \pi_{0} }
\end{equation}

\begin{equation}
\frac{\Pr[0_{2} | ``No"]} {\Pr[0_{2} | ``Yes"]}  = \frac{ \pi_{0} }{ \pi_{0_{2}} + \pi_{s} }
\end{equation}

\begin{equation}
\label{eq:binaryleakage}
\epsilon_{DP} = \max \Bigg(\ln \bigg(\frac{ \pi_{0} + \pi_{s}}{ \pi_{0} }\bigg),\ln \bigg(\frac{ \pi_{0} }{ \pi_{0} + \pi_{s} } \bigg) \Bigg) 
\end{equation}

For \textbf{output 1, round 2}  ($1_{2}$) the privacy leakage is as follows:

\begin{equation}
\epsilon_{DP} = \max \Bigg({\ln \bigg(\frac{\Pr[1_{2} | ``Yes"]} {\Pr[1_{2} | ``No"]} \bigg),\ln \bigg(\frac{\Pr[1_{2} | ``No"]} {\Pr[1_{2} | ``Yes"]} \bigg)} \Bigg)
\end{equation}

\begin{equation}
\frac{\Pr[1_{2} | ``Yes"]} {\Pr[1_{2} | ``No"]} = \frac{ 1 - \pi_{0} }{ 1 - \pi_{0} - \pi_{s} }
\end{equation}

\begin{equation}
\frac{\Pr[1_{2} | ``No"]} {\Pr[1_{2} | ``Yes"]}  = \frac{ 1 - \pi_{0} - \pi_{s} }{ 1 - \pi_{0} }
\end{equation}

\begin{equation}
\label{eq:binaryleakage}
\epsilon_{DP} = \max \Bigg(\ln \bigg(\frac{ 1 - \pi_{0} }{ 1 - \pi_{0} - \pi_{s} }\bigg),\ln \bigg(\frac{ 1 - \pi_{0} - \pi_{s} }{ 1 - \pi_{0} } \bigg) \Bigg) 
\end{equation}

~ \\
\noindent \textbf{(Multiple Values)} The differential privacy leakage is measured as the maximum ratio of the multiple output given the underlying truthful answer is any combination of two values of the output of size $V$.

In round one, there is no privacy leakage as every output is indistinguishable and equally likely given the truthful answer is any of the output. In round two, for any two outputs whereby a data owner would not truthfully output these values, the outputs are indistinguishable and there is no privacy leakage. 

The only privacy leakage occurs when the data owner truthfully responds for \textbf{output V', round two}. Without loss of generality, let the data owners truthful output be V' which can take on any value of the output from $1...V$. Let a data owner's output which is \textit{not} their truthful value is $\neg V'$ as opposed to their truthful value of $V'$.

\begin{equation}
\epsilon_{DP} = \max \Bigg({\ln \bigg(\frac{\Pr[V'_{2} | V']} {\Pr[V'_{2} | \neg V']} \bigg),\ln \bigg(\frac{\Pr[V' | \neg V']} {\Pr[V' | V']} \bigg)} \Bigg)
\end{equation}

\begin{equation}
\frac{\Pr[V'_{2} | V']} {\Pr[V'_{2} | \neg V']} = \frac{ \pi_{V'} + \pi_{s}}{ \pi_{V'} }
\end{equation}

\begin{equation}
\frac{\Pr[V'_{2} | \neg V']} {\Pr[V'_{2} | V']}  = \frac{ \pi_{V'} }{ \pi_{V'} + \pi_{s} }
\end{equation}

\begin{equation}
\label{eq:multipleleakage}
\epsilon_{DP} = \max \Bigg(\ln \bigg( \frac{ \pi_{V'} + \pi_{s}}{ \pi_{V'} } \bigg),\ln \bigg( \frac{ \pi_{V'} }{ \pi_{V'} + \pi_{s} } \bigg) \Bigg) 
\end{equation}

The additional output of 0 for the multiple value scenario reduces the leakage opportunities as compared to the binary value scenario.
 \section{Evaluation}

Our evaluation examines the question of accuracy and privacy leakage. How well does the \titlename mechanism estimation error scale as the population that truthfully responds ``No" increases? How well does the \titlename mechanism compare to the Randomized Response mechanism for both accuracy and privacy leakge?

\subsection{Accuracy}
\label{sec:evaluation:accuracy}

\begin{figure*}
\begin{minipage}{.32\textwidth}
        \centering
    \includegraphics[width=1\columnwidth]{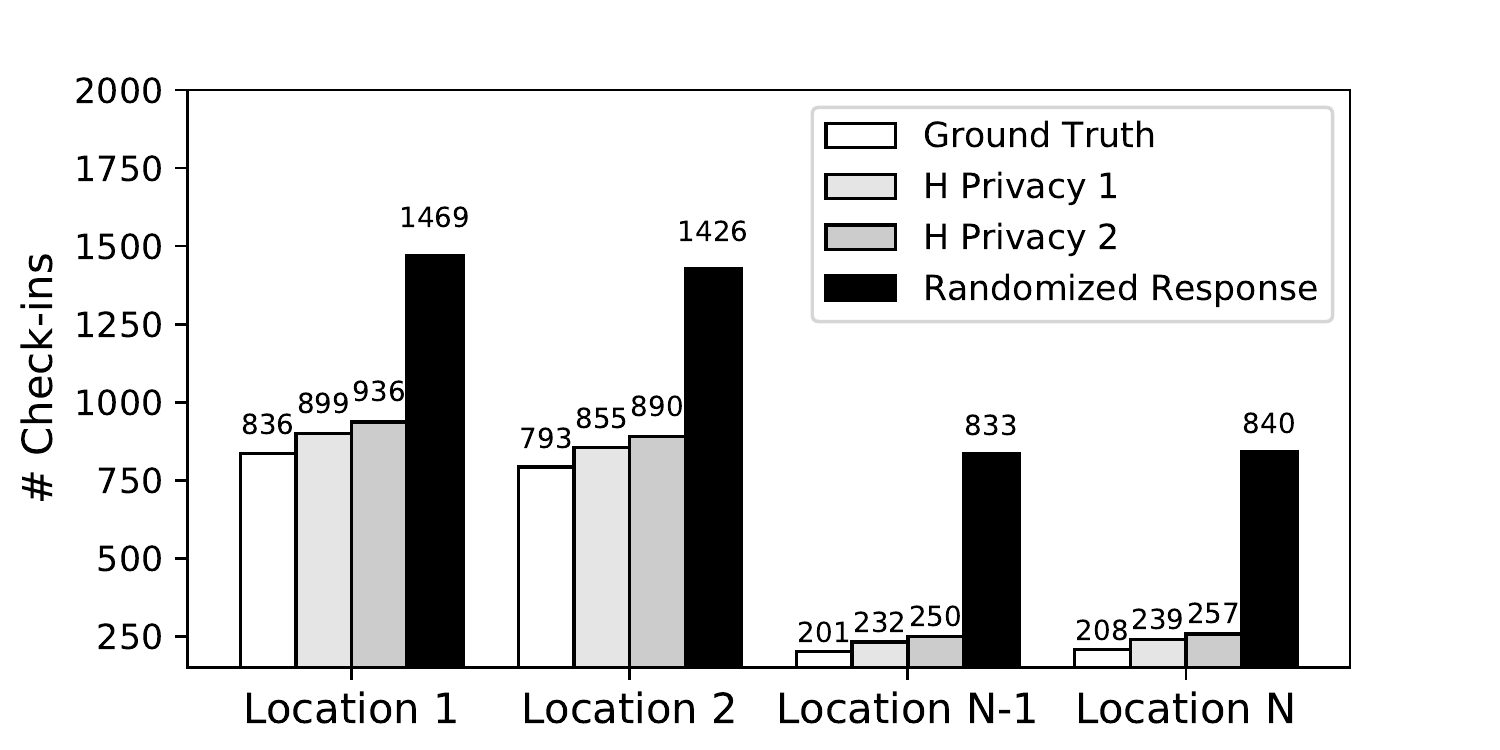}
    \caption{\textbf{(Gowalla)} Check-in counts out of 1 million total check-ins for each distinct location.}
    \label{fig:gowalla}
\end{minipage}\hfill
\begin{minipage}{.32\textwidth}
    \centering
    \includegraphics[width=1\columnwidth]{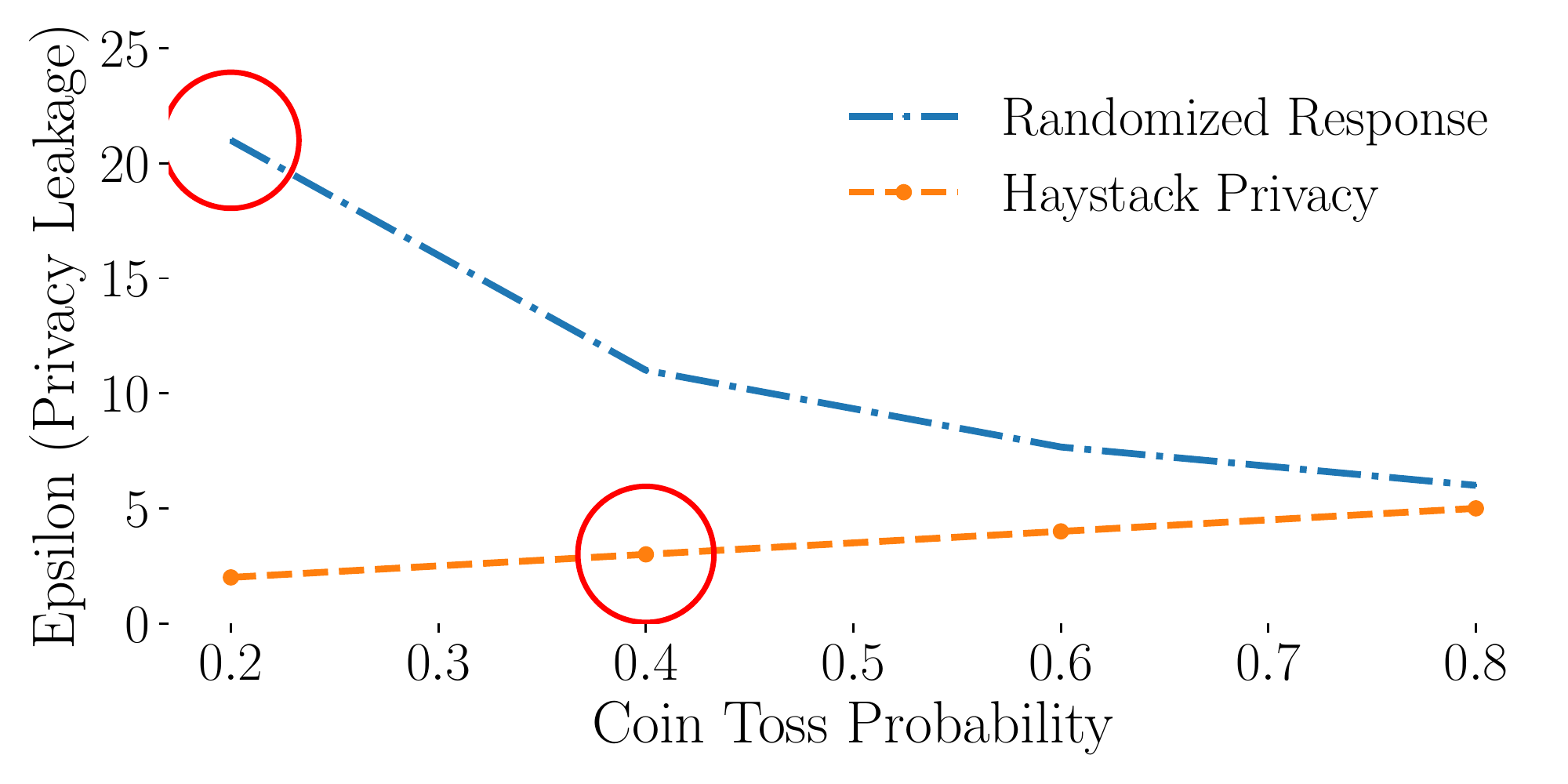}
    \caption{\textbf{(Privacy Leakage)} Privacy leakage corresponding to the coin toss heads success probability. Higher epsilon leaks more information. The red circles indicate the evaluation parameters used.}
    \label{fig:epsiloncomparison}
\end{minipage}\hfill
\begin{minipage}{.32\textwidth}
    \centering
    \includegraphics[width=1\columnwidth]{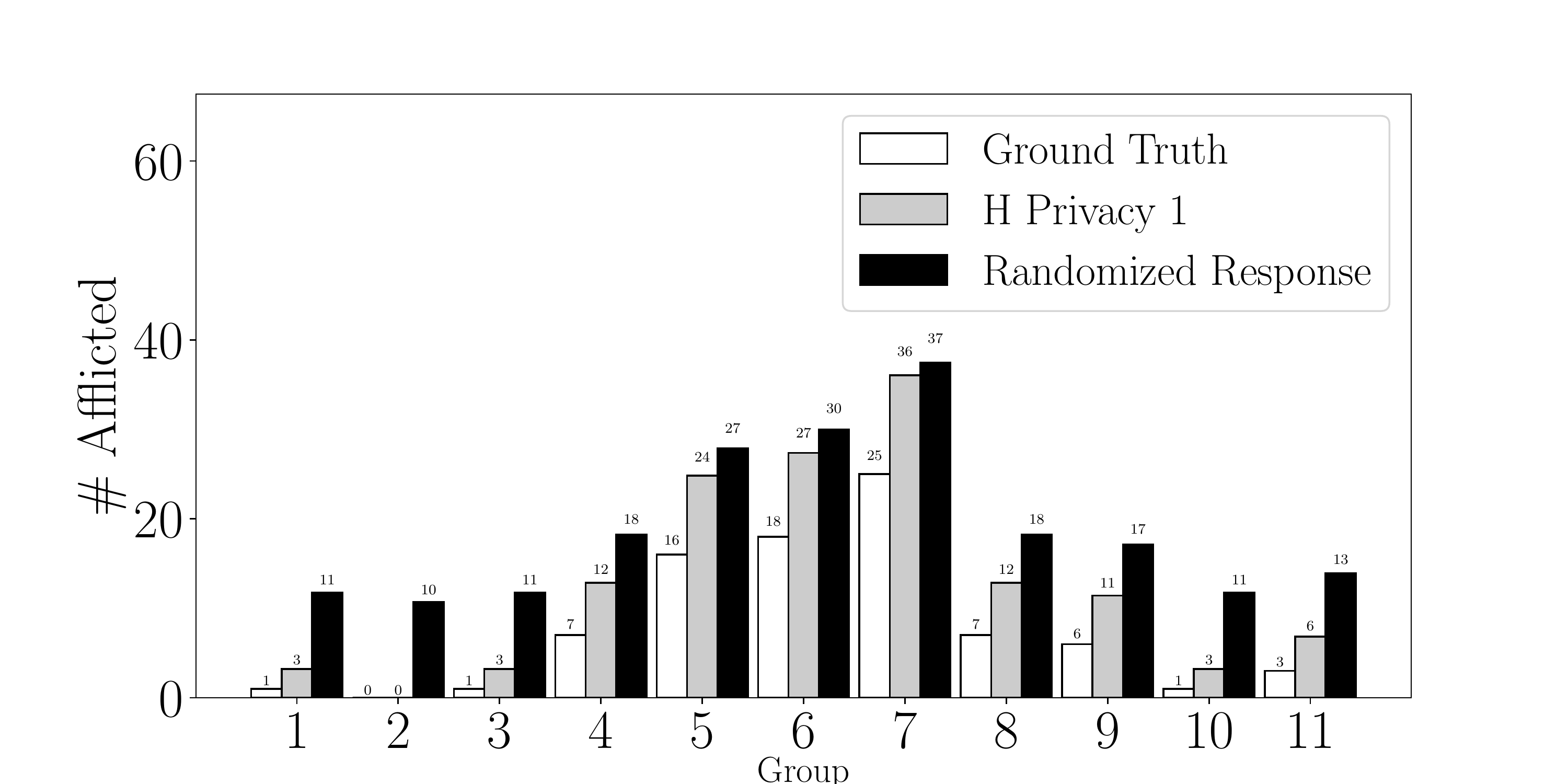}
    \caption{\textbf{(Breast Cancer)}  Number of afflicted individuals out of 286 grouped by tumor size. }
    \label{fig:brst-tsize}
\end{minipage}
\end{figure*}

\begin{figure*}
\begin{minipage}{.32\textwidth}
    \centering
    \includegraphics[width=1\columnwidth]{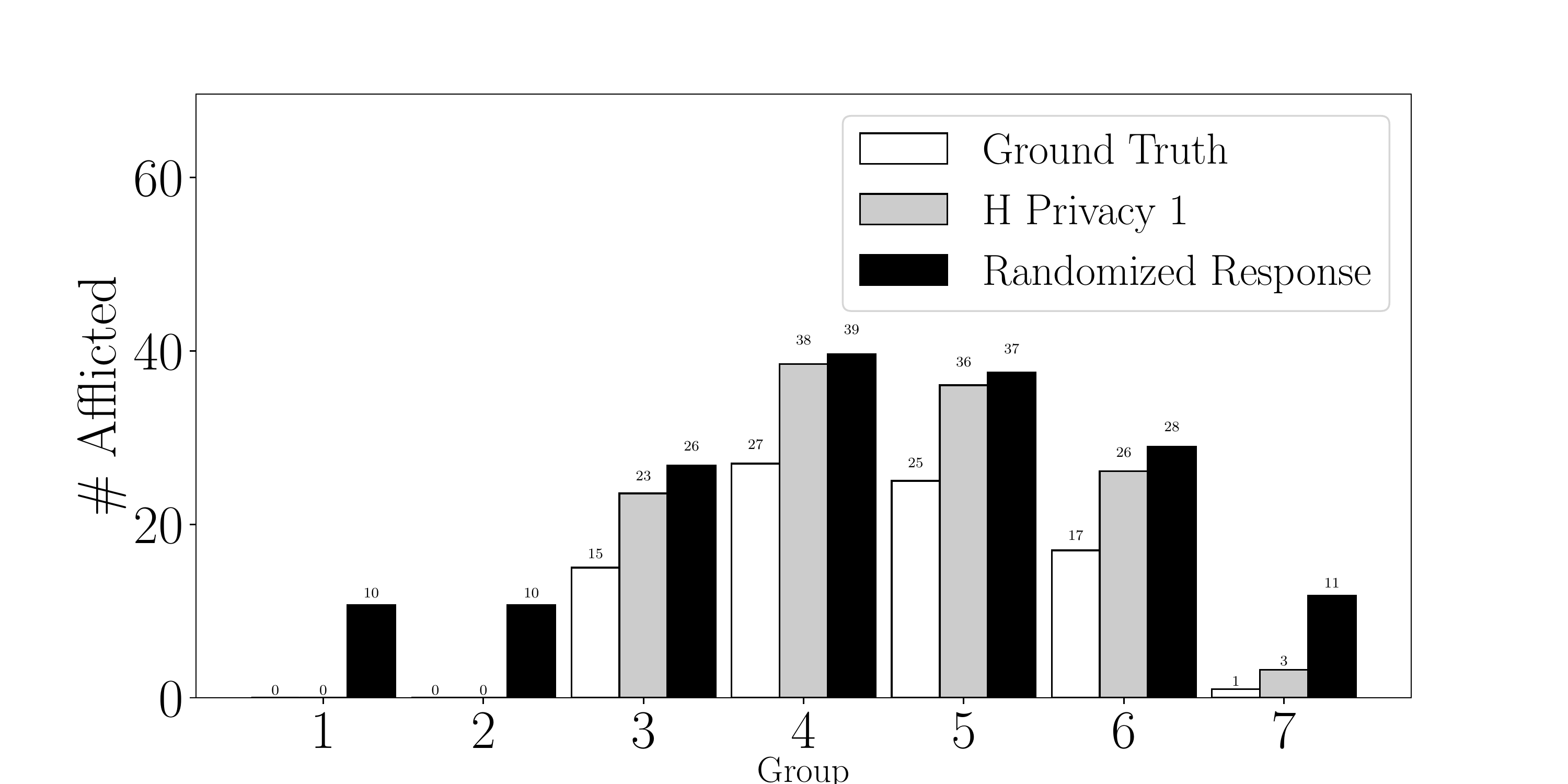}
    \caption{\textbf{(Breast Cancer)}  Number of afflicted individuals out of 286 grouped by age range. }
    \label{fig:brst-age}
\end{minipage}\hfill
\begin{minipage}{.32\textwidth}
    \centering
    \includegraphics[width=1\columnwidth]{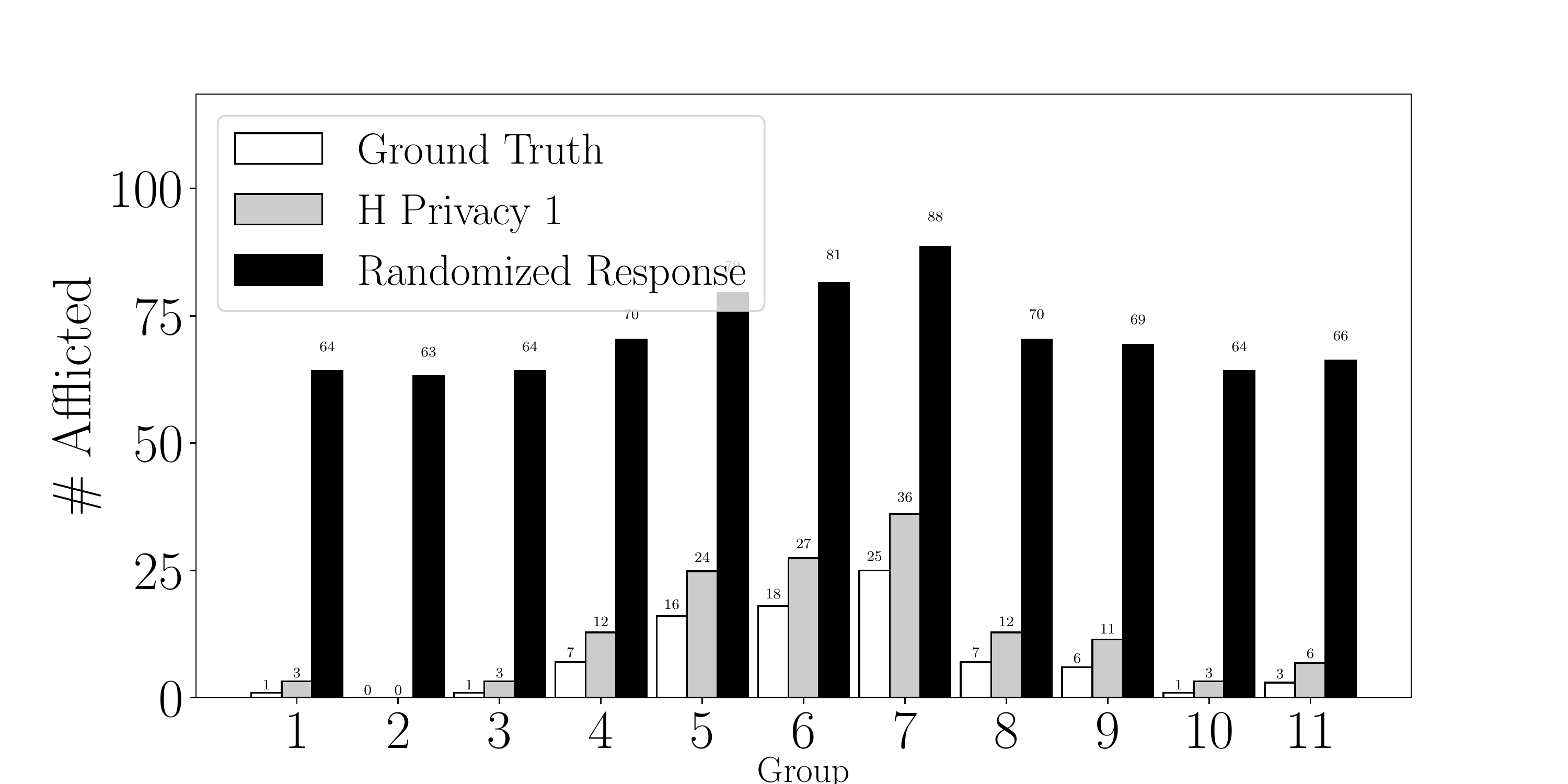}
    \caption{\textbf{(Breast Cancer)}  Number of afflicted individuals out of 10,000 grouped by tumor size.}
    \label{fig:brst-tsize-scale}
\end{minipage}\hfill
\begin{minipage}{0.32\textwidth}
    \centering
    \includegraphics[width=1\columnwidth]{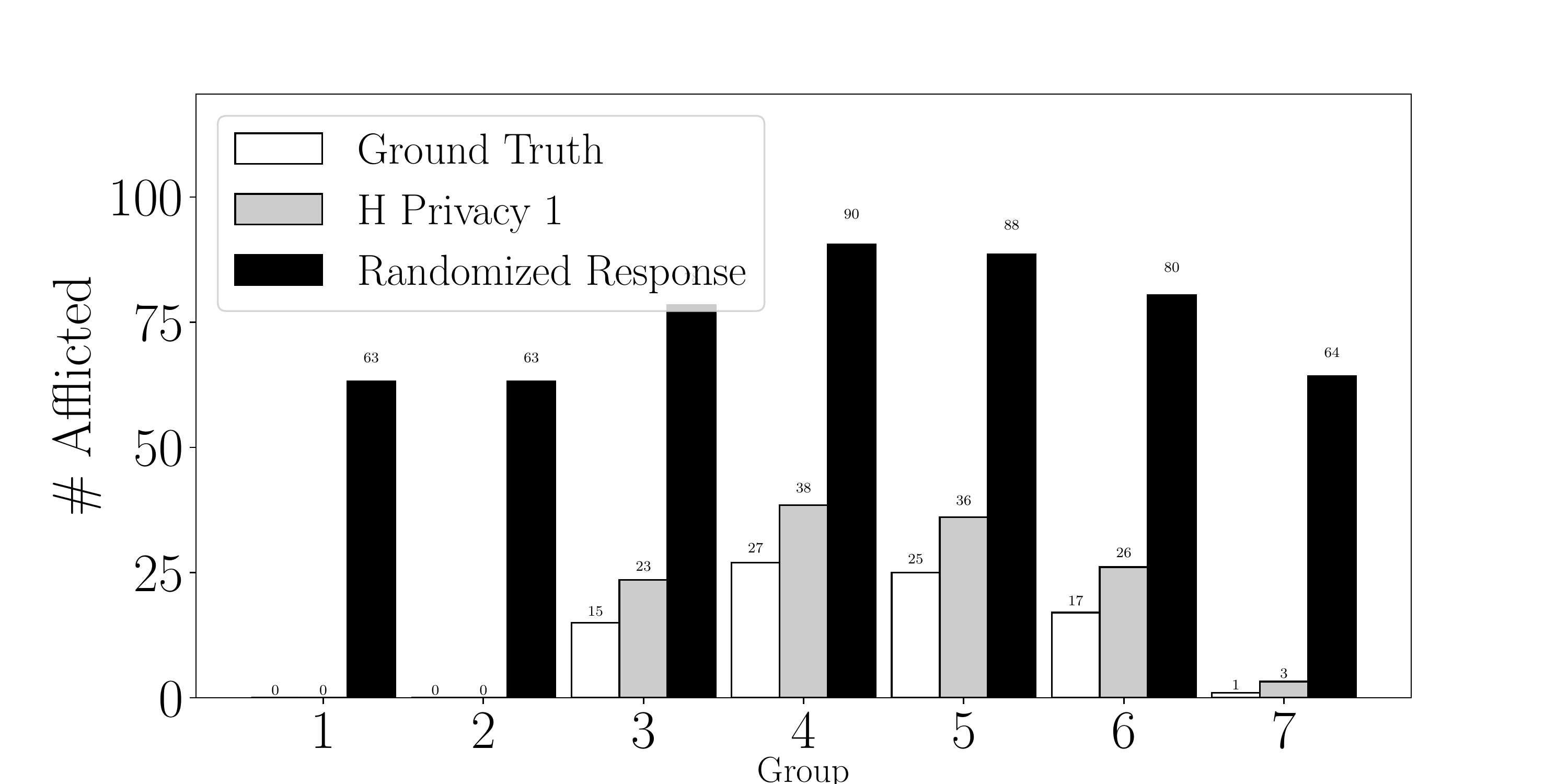}
    \caption{\textbf{(Breast Cancer)}  Number of afflicted individuals out of 10,000 grouped by age range.}
    \label{fig:brst-age-scale}
\end{minipage}
\end{figure*}

~ \\
\noindent \textbf{(Gowalla)} We evaluate over the Gowalla dataset~\cite{DBLP:conf/kdd/ChoML11,snapnets}. Gowalla was a location-based social network where users check-in to popular places (e.g., restaurants, cafes, tourist attractions) and establish social networks connections.  Our evaluation is over New York City. Each checkin contains a location coordinate (latitude, longitude) and location ID which maps to a unique check-in location. For our purposes we use the location ID as specified by Gowalla.

Figure~\ref{fig:gowalla} compares the \titlename mechanism with the Randomized Response and the ground truth using the Gowalla dataset. We show four distinct check-in locations  (rather than all locations for ease of illustration). There are over one million check-ins across all the locations.  H Privacy 1 has a sampling parameter of 45\% and H Privacy 2 has a sampling parameter of 25\%. The Randomized Response mechanism has parameters $flip_1=0.8$ and $flip_2=0.2$. Each bar in the graph represents the absolute error from the ground truth with a 95\% confidence bound. 

The randomized response mechanism has from 1.6 to 3.5 times higher worst case error bound as compared to the \titlename mechanism. The Randomized Response sampling error is due to the second coin toss sampling error as described in Section~\ref{sec:samplingnoise}. \titlename is able to maintain a significantly lower error bound due to the estimation calibration in each round as described in Section~\ref{sec:haystackmechanism}.

~ \\
\noindent \textbf{(Breast Cancer)} Next, we evaluate over the Breast Cancer Dataset~\cite{breast-cancer-dataset,Lichman:2013}. The dataset contains 286 breast cancer patient attributes (10 total attributes such as age, tumor size, menopause). For our evaluation we select the attributes of recurrence, age, and tumor size. The dataset publishes these attributes as ranges. That is, ages are grouped $``10-19",``20-29",...,``90-99"$. Tumor sizes are grouped  $``0-4",``5-9",....,``55-59"$. Each group is assigned an integer identifier. The integers are assigned in increasing order (i.e., $``10-19"$ is group 1, $``20-29"$ is group 2, etc). H Privacy 1 has a sampling parameter of 45\%. The Randomized Response mechanism has parameters $flip_1=0.8$ and $flip_2=0.2$.

Figure~\ref{fig:brst-tsize} and Figure~\ref{fig:brst-age} evaluates the absolute error with a 95\% confidence bound whereby we privatize only the \textit{attributes} of each patient, such as age and tumor size. However, an adversary that knows that a particular patient was involved in the published data set knows definitively that the particular data owner has breast cancer. Thus, we need to increase the non-cancer population.

Figure~\ref{fig:brst-tsize-scale} and Figure~\ref{fig:brst-age-scale} evaluates the absolute error with a 95\% confidence bound whereby we \textit{increase} the query population from 286 to 10,000 data owners (the number of non-cancer data owners is 9714). The Randomized Response quickly grows in absolute error due to the sampling error while \titlename maintains constant error.

\subsection{Privacy}
\label{sec:evaluation:privacy}

We now examine the privacy leakage of the \titlename mechanism. Figure~\ref{fig:epsiloncomparison} evaluates the privacy leakage comparing the \titlename mechanism and the Randomized Response mechanism. \titlename uses the privacy guarantee equation defined in Equation~\ref{eq:multipleleakage} to measure the privacy leakage. The privacy guarantee equation for the Randomized Response mechanism privacy leakage details can be found in Appendix~\ref{sec:randomizedresponseprivacy}.

The coin toss parameters used in Figure~\ref{fig:epsiloncomparison} has Randomized Response $flip1=0.8$. The x-axis varies the second coin toss heads success probability $\pi_2$. The privacy leakage decreases as the privacy noise is increased (by increasing the second coin toss head success probability). However, more privacy noise is added as the cost of accuracy.

The \titlename mechanism uses a sampling parameter of $0.45$. As we will see we will maintain lower privacy leakage as compared to the randomized response mechanism. The x-axis refers to the con toss heads success probability for the truthful response $\pi_{\perp_{V'}}$. Increasing this coin toss heads success probability increases the amount of privacy leakage as the distance of the honest responses and privatized responses grows larger.

The red circles correspond to the coin toss head success probabilities used in the dataset evaluation in Section~\ref{sec:evaluation:accuracy}. We could increase the randomized response second coin toss $\pi_2$ probability to decrease the privacy leakage at the cost of increased absolute error.

 \section{Conclusion}

In this paper, we introduce the \titlename Mechanism and show how to achieve differential privacy in the distributed setting by sampling alone. We show that we can maintain constant error (i.e., absolute error from the ground truth) even as the population increases and achieve a as much as four times lower privacy leakage as compared to randomized response. 
\appendix

\section{Randomized Response Privacy Guarantee}
\label{sec:randomizedresponseprivacy}

\subsubsection{Privacy Guarantee of Randomized Response}
\label{sec:priv_g}
The randomized response mechanism achieves $\epsilon$-differential privacy, where:

\small
\begin{multline*}
\epsilon = \max \bigg({\ln \Big(\frac{\Pr[\textrm{Resp=`Yes' \textbar `Yes'}]} {\Pr[\textrm{Resp=`Yes' \textbar `No'}]} \Big), \ln \Big(\frac{\Pr[\textrm{Resp=`Yes' \textbar `No'}]} {\Pr[\textrm{Resp=`Yes' \textbar `Yes'}]}\Big)} \bigg)
\end{multline*}
\normalsize

More specifically, the randomized response mechanism~\cite{fox1986randomized} achieves $\epsilon$-differential privacy, where:
\begin{equation}
\label{eqn:e-forced}
\epsilon = \ln \Big( \frac{\pi_1 + (1 - \pi_1) \times \pi_2}{(1 - \pi_1) \times \pi_2} \Big)
\end{equation}

That is, if a data owner has the sensitive attribute $A$, then the randomized answer will be ``Yes'' with the probability of `$\pi_1 + (1 - \pi_1) \times \pi_2$'.  Else, if a data owner does not have the sensitive attribute, then the randomized answer will become ``Yes'' with the probability of `$(1 - \pi_1) \times \pi_2$'.
 
\bibliographystyle{IEEEtran}
\bibliography{bib/fss,bib/privacy,bib/vehicles,bib/mpc,bib/learning,bib/streamprivacy}

\end{document}